\documentclass[conference]{IEEEtran}
\IEEEoverridecommandlockouts
\usepackage{cite}
\usepackage{amsmath,amssymb,amsfonts}
\usepackage{graphicx}
\usepackage{booktabs}
\usepackage{multirow}
\usepackage[english]{babel}

\def\BibTeX{{\rm B\kern-.05em{\sc i\kern-.025em b}\kern-.08em
    T\kern-.1667em\lower.7ex\hbox{E}\kern-.125emX}}
\begin{document}

\title{Improving THz Quality-of-Transmission with Systematic RLNC and Auxiliary Channels}
\author{\IEEEauthorblockN{Cao Vien Phung, Anna Engelmann, Thomas Kuerner and Admela Jukan}
\IEEEauthorblockA{Technische Universit\"at Braunschweig, Germany\\
Email: \{c.phung, a.engelmann, t.kuerner and a.jukan\}@tu-bs.de
}}
\maketitle

\begin{abstract}
In this paper, we propose a novel solution that can improve the quality of THz transmission with systematic random linear network coding (sRLNC) and a low-bitrate auxiliary channel. To minimize complexity of channel coding, we complement a generic low complexity FEC code by a low complexity sRLNC. To increase the overall throughput of THz transmission, we propose to send the native data and coding redundancy in parallel over $2$ differently configured THz channels, i.e., over $1$ high bit rate main channel and $1$ low bit rate low error rate auxiliary channel. The results show, that the main THz channel supported by low bit rate auxiliary channel can use a higher level modulation format and sent over longer distances with a higher throughput.
\end{abstract}

\section{Introduction}

The terahertz (THz) frequency band ranging from $0.3$ THz to $10$ THz is the prime candidate to fulfilling the capacity needs of the exponentially growing data volumes in wireless networks. The dynamic configuration of the THz channel is still a technology challenge, as it strongly depends on THz system design and the environment. An increasing signal modulation level and THz transmission distance (even when ranging in meters) can significantly decrease a Signal-to-Noise ratio (SNR) resulting in a high bit error rate (BER) of transmitted data. At the same time, both SNR and BER have a limited configuration flexibility, while modulation level and transmission distance need to be set up as high as possible to provide both a broad THz coverage and a high transmission rate in the terahertz range. Additional decrease in SNR is also often caused by molecular absorption in the atmosphere \cite{7321055, 7444891}. To solve the problem of low SNR of THz signal when sending over longer transmission distances with high modulation format, and under atmospheric effects, the research community is in search for novel ideas and solutions. 

So far, the key components studied for high-speed THz communication with compensated path attenuation were fast and efficient amplitude and phase modulators, high gain and massive multiple-input multiple-output (MIMO) antennas and the efficient THz beamforming \cite{Zhou:17, Rout:2016, Mittendorff:2017,Khalid:2016, Ullah:2019, Jiang:2018, Lin:2015, Yan:2019}. In \cite{7444891}, a WiFi channel is used together with the THz channel to estimate the distance to the THz receiver and to measure the relative air humidity. The research on an adequate channel coding method is however still in its infancy. Currently,  the state-of-the-art Forward Error Correction (FEC) Codes such as Reed Solomon, Low Density Parity Check (LDPC) are under discussion \cite{wehn_norbert_2018_1346686, wehn_norbert_2019_3360520}. These codes are known for a high throughput, but also high computational complexity. The Polar code, which is under consideration as well, has lower complexity, but cannot provide a high throughput due to amount of coding redundancy required. Thus, implementing low-complexity codes without reducing the THz channel throughput is an open challenge.

In this paper, we propose a novel solution for THz transmission that can tackle the challenges of coding complexity and the throughput. We first propose a low-complexity systematic random linear network coding (sRLNC), a coding technique presented in \cite{5513768}, in addition to, and to complement a generic low complexity FEC \cite{Wang:2006}. sRLNC is a good choice as it does not change the source data during encoding process in sender which additionally reduces decoding overhead. During encoding process, a coding redundancy is generated for erasure error correction at the receiver. Since the overall throughput of THz transmission directly relates to the amount of coding redundancy required, we furthermore propose to send the source data and coding redundancy in parallel over two differently configured channels, i.e., main and auxiliary. As a result, the source data is transmitted over a high bit rate main THz channel and coding redundancy is sent over a low bit rate low error rate auxiliary channel, i.e., not necessarily THz. We investigate different modulation formats and transmission distances, and, with our analytical model, evaluate the amount of sRLNC redundancy, the related code rate and the transmission rate of auxiliary channel required to ensure reliable THz transmission. The results show that, when supported by low bit rate auxiliary channel, the main THz channel can use a higher level modulation format and reach longer distances with higher throughput.

The remainder of this paper is organized as follows. In Sec. \ref{schemes}, we design THz transmission system with sRLNC. Sec. \ref{analysis} gives an analysis for evaluation of amount of redundancy. Sec. \ref{Numericalresults} shows numerical results.  Sec. V presents a discussion on system implementation. Sec. \ref{conc} concludes the paper.

\begin{figure*}[!ht]
\centering
\includegraphics[width=2\columnwidth]{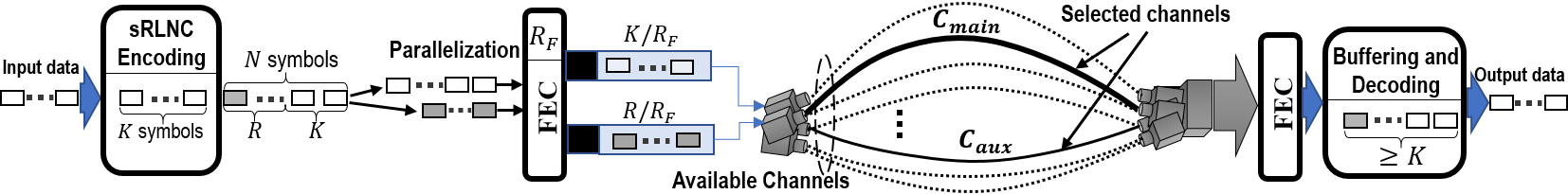}
  \vspace{-0.1cm}
  \caption{Multi-Channel THz transmission system with systematic random linear network coding (sRLNC).}
  \vspace{-0.6cm}
  \label{codingscenario}
\end{figure*}

\section{System design}\label{schemes}
In this section, we present a THz transmission system in configuration with two channels and sRLNC, based on \cite{patent} considering THz channel characteristics presented in \cite{7444891}. The notations are summarized in Table \ref{tab:table1}.

Figure \ref{codingscenario} presents a reference architecture of THz transmission, where coded data from the source are divided and simultaneously transmitted over $2$ THz channels; in general more than two channels can be used, but in this paper we limit the discussion to two channels for clarity. The source performs sRLNC encoding and parallelization of incoming source data prior to THz transmission. As pre-processing, any input bit stream is split into \emph{symbols} of fixed size, $s$ bits each, referred to as \emph{native symbols}.

We refer to $K$ native symbols encoded together as a \emph{generation}, where $K$ is a fixed generation size. The number of generations per input data depends on the amount of input symbols. For simplicity, Fig. \ref{codingscenario} only shows one generation with $K$ native symbols. The sRLNC coding process is performed over a finite field $\mathbb{F}_{2^s}$, where $s$ is the symbol size and coding coefficients can be randomly selected. During encoding, $R$ redundant symbols are generated from $K$ native symbols. Although each redundant symbol presents a linear combination of the $K$ native symbols combined with coding coefficients, the native symbols keep unchanged after encoding and the size of any native and any redundant symbols keeps $s$ bits. Any generation has a fixed and constant number of native symbols, whereby the number of redundant symbols is adapted to the expected bit error rate (BER) on the related main channel.

After sRLNC encoding, $N=K+R$ symbols build a so-called \emph{coded generation}, which is parallelized and distributed over $2$ channels, i.e., one high bit rate main THz channel and one low bit rate auxiliary channel. We refer to the parallelization process as separation and distribution of the native symbols and redundancy over two channels. After the parallelization process, a FEC per channel is required. In this paper, we do not consider any specific FEC in the physical layer, and apply sRLNC in higher layers (e.g., Ethernet) to complement FOR any FEC mechanisms, i.e, when FEC with a code rate $R_F$ either fails or yields insufficient performance. Thus, before the THz signal transmission, $K$ native symbols and its related redundancy are packetized and protected by FEC. As a result, the generation size and its redundancy are increased as $\frac{s \cdot K}{R_F}$ bits and $\frac{s \cdot R}{R_F}$ bits, respectively.

Finally, $K$ native symbols are sent on the main channel, while $R$ redundant symbols are transferred on the auxiliary channel. The proposed THz system utilizes a main channel with the high-level modulation format to reach a high bit rate. In contrast, the related low bit rate auxiliary channel is used to reliably deliver coding redundancy for erasure correction of symbol errors occurred on the main channel. As a result, the main and the related auxiliary THz channel need to be perfectly tuned to each other regarding modulation level, transmission distance, signal power and other parameters to be able to deal with bit errors and symbol lost occurred during THz transmission. This presents a significant engineering challenge.

At the receiver, as it is commonly the case, FEC will be performed first, whereby the symbols with bit errors will be removed. In this scenarios, sRLNC redundancy from the auxiliary channel is used to compensate those symbol losses, whereby any redundant symbol can replace any native symbol from the same coded generation. The decoding process is performed by running Gauss-Jordan elimination. The decoding and, thus, correct data reception is only successful, when at least $K$ out of $K+R$ symbols from any coded generation arrive at decoder. To this end, buffering at the receiver side is necessary to temporarily store symbols until any $K$ out of $N$ symbols arrive and decoding process can start. The buffering process in a high speed system like this presents another significant engineering challenge.

\begin{table}[t!]
  \centering
  \caption{List of notations.}
  \label{tab:table1}
  \begin{tabular}{ll}
    \toprule
    Notation & Meaning\\
    \midrule
    $P_b$ & Residual bit error rate on the main channel.\\
    $P_s$ & Residual symbol error rate on the main channel.\\
    $p_e$ & Expected BER on the main channel.\\
    $s$ & Symbol size in bits.\\
    $R_F$ & Code rate of FEC code.\\
    $R_L$ & Code rate of sRLNC.\\
    $K$ & Generation size in symbols. \\
    $R$ & Redundancy in symbols. \\
    $N$ & Size of coded generation in symbols.\\
    $C_{main}$ & Transmission rate of the main channel.\\
    $C_{aux}$ & Transmission rate of the auxiliary channel.\\
    $\tau_{main}$ & Propagation delay on the main channel.\\
    $\tau_{aux}$ & Propagation delay on the auxiliary channel.\\
    $t_{main}$ & Transmission delay on the main channel.\\
    $t_{aux}$ & Transmission delay on the auxiliary channel.\\
    $T_{main}$ & Total delay on the main channel.\\
    $T_{aux}$ & Total delay on the auxiliary channel.\\
    $d_{main}$ & Transmission distance of the main channel.\\
    $d_{aux}$ & Transmission distance of the auxiliary channel.\\
     $c_p$ & Propagation speed of light, $c_p=3\cdot 10^{8}$m/s.\\
    \bottomrule
  \end{tabular}\vspace{-0.5cm}
\end{table}
\section{Analysis}\label{analysis}
In this section, we provide an analytical model for the code rate evaluation and calculation of minimal transmission rate of auxiliary channel for the proposed THz transmission system. Without loss of generality, we assume a generic FEC code with a code rate $R_F$ and the minimal Hamming distance $\Delta^K_{min}=\frac{K \cdot s}{R_F}-(K \cdot s)$ and $\Delta^R_{min}=\frac{R \cdot s}{R_F}-(R \cdot s)$. Since we assume that redundancy sent over auxiliary channel is negligibly affected by bit errors, which can be corrected during FEC process, we further assume that  the residual BER of the auxiliary channel is $0$. Thus, to derive residual BER $P_b$ on the main channel after FEC process at receiver, we consider the minimal Hamming distance defined as follows
\begin{equation}\label{ham}
\Delta_{min} =\frac{K \cdot s}{R_F}-(K \cdot s) \equiv  \Delta^K_{min}
 \end{equation} 
For the main channel, the number of erroneous bits that can be detected by a FEC code is determined as $t_e=\Delta_{min}-1$. The number of corrected erroneous bits is $t_k=\frac{\Delta_{min}-2}{2}$ or $t_k=\frac{\Delta_{min}-1}{2}$, if $\Delta_{min}$ is even or odd, respectively. Then, the expected residual bit error rate $P_b$ of the main channel is
\begin{equation}\label{eqBER}
 P_b = \frac{(K \cdot s \cdot p_e) - (R_F \cdot t_k)}{K \cdot s},
 \end{equation} 
where $p_e$ is the expected BER on the main transmission channel. Note that in the case of $P_b \leq  0$ means that all erroneous bits on the main channel are corrected by FEC. Some of bit errors can not be corrected by FEC resulting in symbol errors described by residual symbol error rate $P_s$, which is can be calculated with Eq. \eqref{eqBER} as
 \begin{equation}
P_s=1-(1-P_b)^{s}
 \end{equation}
Then, the expected number of erroneous symbols per generation is $P_s\cdot K$. To ensure successful recovery of source data after THz transmission, at least $P_s \cdot K$ additional coded symbols need to be sent over the auxiliary channel, i.e., 
 \begin{equation}\label{redund}
R\geq P_s\cdot K,
  \end{equation}  
where the amount of redundancy $R$ sent on the auxiliary channel is a function of the residual symbol error rate on the main channel, while generation size $K$ is constant. 

In contrast to the constant code rate $R_F$ of the generic FEC code, the overall code rate of sRLNC is a function of channels' configuration and the amount of coding redundancy required to protect one generation. Since the total number of the native and redundant symbols sent to THz receiver is determined as $N=K+R$, where  $R\geq 0$ and $N \geq K$, the code rate provided by sRLNC can be calculated as 
\begin{equation}\label{CR}
R_L=\frac{K}{N}
 \end{equation}

For the main channel, $K$ native symbols will arrive at the receiver after time interval $T_{main}=t_{main}+\tau_{main}=\frac{K \cdot s}{R_F \cdot C_{main}} + \frac{d_{main}}{c_p}$, where $t_{main}$ is the transmission delay of one generation with $K$ native symbols, $\tau_{main}$ denotes the propagation delay, $C_{main}$ represents a transmission rate of the main THz channel, $d_{main}$ stands for the transmission distance between sending and receiving THz antennas and $c_p$ is the propagation speed of light. For auxiliary channel, the redundancy of $R$ coded symbols will arrive at the receiver after time interval $T_{aux}=t_{aux}+\tau_{aux}=\frac{R \cdot s}{R_F \cdot C_{aux}} + \frac{d_{aux}}{c_p}$, where $t_{aux}$ is the transmission delay for sending $R$ redundant symbols on auxiliary channel, $\tau_{aux}$ denotes a propagation delay of auxiliary channel of length $d_{aux}$ and $C_{aux}$ represents the transmission rate of auxiliary channel. To avoid very large buffer, the THz system should be configured so that symbols from the main and auxiliary channel arrive at the THz receiver simultaneously, i.e., $T_{main}=T_{aux}$ resulting in $\frac{K \cdot s}{R_F \cdot C_{main}} + \frac{d_{main}}{c_p}=\frac{R \cdot s}{R_F \cdot C_{aux}} + \frac{d_{aux}}{c_p}$. As a result, the transmission rate of auxiliary channel is a function of transmission distance of main and auxiliary channels and can be derived as 
\begin{equation}\label{rateaux}
\begin{split}
C_{aux}&=\frac{R \cdot s}{R_F \cdot [\frac{d_{main}-d_{aux}}{c_p}+\frac{K \cdot s}{R_F \cdot C_{main}}]}=\\&=\frac{R \cdot s\cdot c_p\cdot C_{main}}{R_F \cdot C_{main}(d_{main}-d_{aux})+c_p \cdot K \cdot s}
 \end{split}
 \end{equation}
Generally, $C_{aux} \geq 0$, whereby $C_{aux}=0$ means that it is not necessary to establish an auxiliary channel and to send redundancy. From Eq. \eqref{rateaux}, we can derive a constraint $R_F \cdot C_{main}(d_{main}-d_{aux})+c_p \cdot K \cdot s>0$, then the transmission distance of auxiliary channel has to be limited as \begin{equation}\label{dlimit}
d_{aux} < \frac{K \cdot s \cdot c_p}{R_F \cdot C_{main}} + d_{main}
 \end{equation}
That distance limit for auxiliary channel is only required for earlier sRLNC decoding start, i.e., to maximally reduce the buffer size at receiver.

\section{Numerical results} \label{Numericalresults}

In this section, we show analytical results for the code rate of sRLNC $R_L$ and for the transmission rate of auxiliary path $C_{aux}$ calculated with Eqs. \eqref{CR}, \eqref{redund}, \eqref{eqBER} and \eqref{rateaux}, respectively. The expected bit error rate (BER) $p_e$ of THz channel is collected from \cite{7444891} and is a function of THz channel type, modulation format and distance. We consider two channel types, i.e.,  Channel B ($660-695$ GHz) and channel C ($855-890$ GHz), which were simulated by assuming the raised cosine filter with roll-off factor $0.4$, utilized omni-directional antennas and provided the transmission rate of $C_{main}=25$ GBd/s$=2 \cdot 10^{11} \cdot M$ bps in \cite{7444891}, where $M$ is the amount of bits per symbol for certain modulation level.  For our evaluation, we assume that any auxiliary channel is configured so that residual BER on the auxiliary channel is approximately equal to $0$. The symbol size and the total number of symbols were set to $s=8$ bits and $K=100$, respectively. The generic FEC has the same code rate of $R_F=0.73$ for any THz channel. The transmission distance $d_{main}$ of main channel between sending and receiving antennas is in interval $[200,2000] cm$ and the modulation formats investigated are 16PSK, 8PSK, QPSK and BPSK. The Hamming distance calculated with Eq. \eqref{ham} is $\Delta_{min}=\frac{100 \cdot 8}{0.73}-(100 \cdot 8)=296$ bits for any generation of $K$ symbols. Then, the number of erroneous bits that can be corrected by FEC in each generation sent over the main THz channel is $t_k=\frac{296-2}{2}=147$ bits.

\begin{figure}[t]
\centering
\includegraphics[width=1\columnwidth]{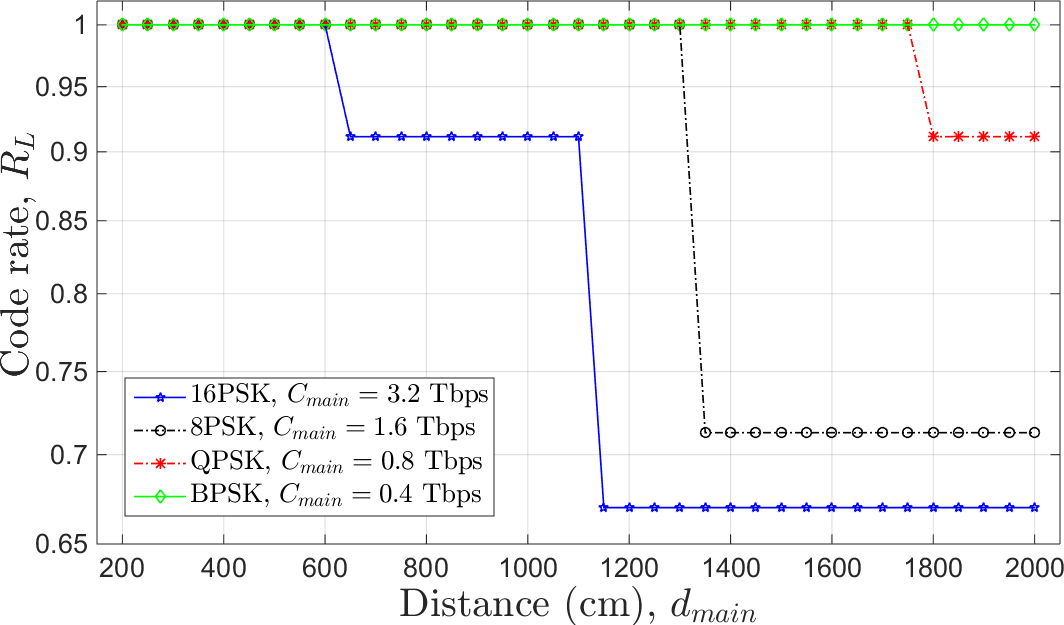}
\caption{Code rate $R_L$ vs. transmission distance of main channel B $d_{main}$.}
\label{2channelB}
\vspace{-0.3cm}
\end{figure}

Fig. \ref{2channelB} presents a code rate of sRLNC for the THz transmission over a high transmission rate THz channel B  supported by any lower transmission rate channel. For BPSK, the sender does not need to send any redundancy, because all bit errors on main channel B can be corrected by FEC resulting in code rate $R_L=1$. In this case, there is no need for an auxiliary channel, however, if the code rate $R_F$ is lower as assumed here, i.e., $R_F<0.73$, the auxiliary channel can be required to provide sRLNC redundancy. For QPSK and 16PSK, there is a need for around $10$ redundant symbols and the resulting code rate $R_L>0.9$ when the distance of main channel $d_{main}$ is $1800 cm$ to $2000 cm$ and $650 cm$ to $1100 cm$, respectively. The modulation format 8PSK only requires redundancy when the distance $d_{main}$ is larger than or equal to $1350 cm$, i.e., $R_L=0.71$. The lowest code rate of $0.67$ is in configuration with 16PSK, when the distance $d_{main}$ is larger than or equal to $1150 cm$.

\begin{figure}[t]
\centering
\includegraphics[width=1\columnwidth]{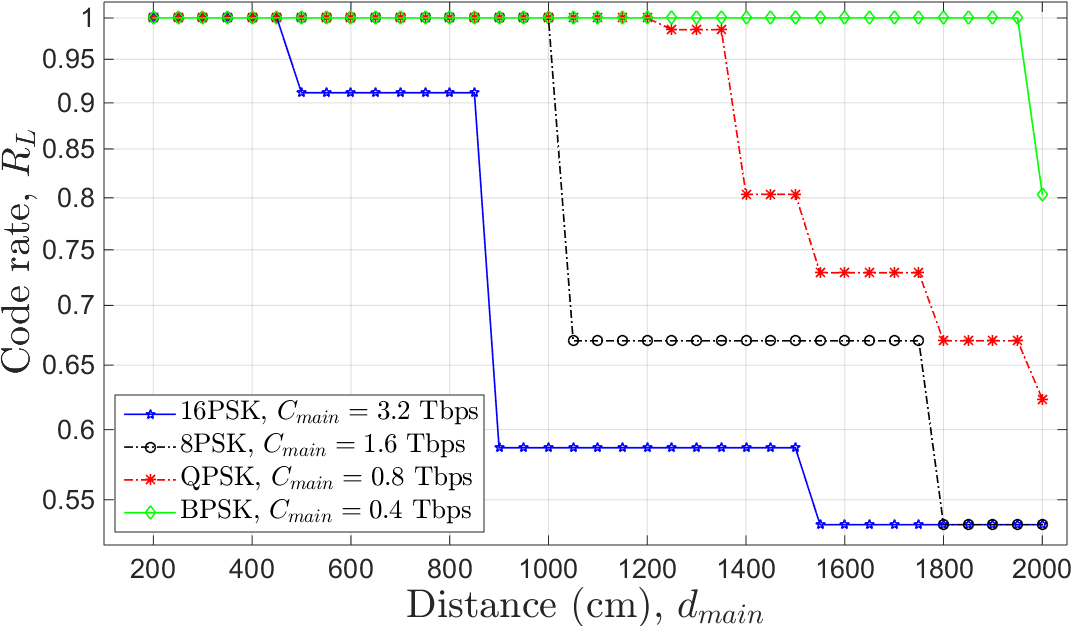}
\caption{Code rate $R_L$ vs. transmission distance of main channel C $d_{main}$.}
\label{2channelC}
\vspace{-0.6cm}
\end{figure}

Figure \ref{2channelC} shows the resulting code rate $R_L$ for THz transmission with sRLNC over a main THz channel configured as channel C and any low transmission rate auxiliary channel. In case of BPSK modulation, the coding redundancy is only required, i.e., $R_L=0.8<1$, when transmission distance of main channel $d_{main}$ is larger than $2000 cm$. Using QPSK, the code rate can vary from $1$ to $0.62$ and decreases with increasing transmission distance $d_{main}$, i.e., from $1200 cm$ to $2000 cm$, respectively. When 8PSK is used and  transmission distance $d_{main}$ is larger than $1000 cm$, there is a need for up to $88$ redundant symbols per generation, which reduce code rate up to $0.53$. For 16PSK, the redundancy is essential when transmission distance $d_{main}$ is larger than or equal to $500cm$, while the code rate $R_L$ is gradually decreased according to increasing distance $d_{main}$ as $0.91$, $0.59$, $0.53$.

Observing Fig. \ref{2channelB} and Fig. \ref{2channelC}, the code rate decreases with an increasing transmission distance and an increasing modulation level.  However, the code rate can be maximal, i.e., $R_L\approx1$ and independent from transmission distance, when BPSK is utilized on main channels. The disadvantage of BPSK modulation format is the resulting low transmission rate. On the other hand, with a code rate $R_L\leq 0.9$, it is possible to use modulation formats up to 16PSK.

Next, we evaluate the transmission rate of auxiliary channel $C_{aux}$ required to support the main THz channel by providing sRLNC redundancy for symbol recovery at THz receiver. The transmission rate of auxiliary channel is calculated with Eq. \eqref{rateaux} as function of coding redundancy sent $R$ and a transmission distance of the main and auxiliary channels, $d_{main}$ and $d_{aux}$. $C_{aux}=0$ means, there is no need for the auxiliary channel and FEC code is able to correct all corrupted bits from the main THz channel.

\begin{figure}[t]
\centering
\includegraphics[width=1\columnwidth]{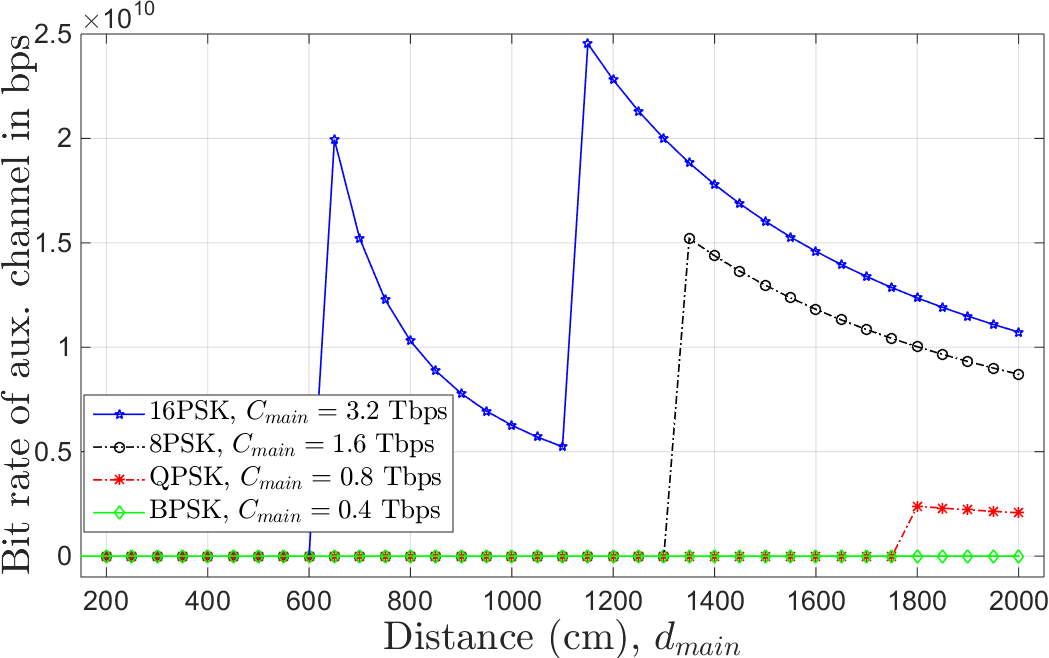}
\caption{Transmission rate of auxiliary channel $C_{aux}$ vs. transmission distance of main channel B, $d_{main}$, when $d_{aux}=500$ cm.}
\label{transratechannelB}
\vspace{-0.3cm}
\end{figure}

First, we fixed the transmission distance of auxiliary channel to $d_{aux}=500cm=5m$ and investigated the transmission rate $C_{aux}$ for all transmission distances and modulation formats of main THz channel.
Fig. \ref{transratechannelB} presents the transmission rate of auxiliary channel, when channel B is utilized as main channel. The sender does not need to use the auxiliary channel for BPSK. $C_{aux}$ gradually decreases from $2.383 \cdot 10^9$ bps to $2.074 \cdot 10^9$ bps for QPSK and from $1.521 \cdot 10^{10}$ bps to $0.871 \cdot 10^{10}$ bps for 8PSK when the transmission distance $d_{main}$ increases. This behavior can be explained by considering amount of redundancy sent, which is  constant, i.e., $41$ and $10$ redundant symbols, for transmission distance $d_{main}$ from intervals [$1350cm$,$2000cm$] and [$1800cm,2000cm$] for 8PSK and QPSK, respectively. At the same time, the propagation delay $\tau_{main}$ of main channel increases with transmission distance $d_{main}$ resulting in a decreasing transmission rate $C_{aux}$ of auxiliary channel, which has a fixed transmission distance $d_{aux}=5m<<d_{main}$. For 16PSK, $C_{aux}$ gradually decreases from $1.994 \cdot 10^{10}$ bps to $0.524 \cdot 10^{10}$ bps, achieves a peak at $2.453 \cdot 10^{10}$ bps, and then keeps decreasing from $2.453 \cdot 10^{10}$ bps to $1.073 \cdot 10^{10}$ bps. The gradual decreasing of $C_{aux}$ can be similarly explained as the case of QPSK and 8PSK and the reason of such a sharp increase is  the increasing amount of additional redundant symbols required, i.e., from $10$ redundant symbols at $d_{main}=1100cm$ to $50$ redundant symbols at $d_{main}=1150cm$.

\begin{figure}[t]
\centering
\includegraphics[width=1\columnwidth]{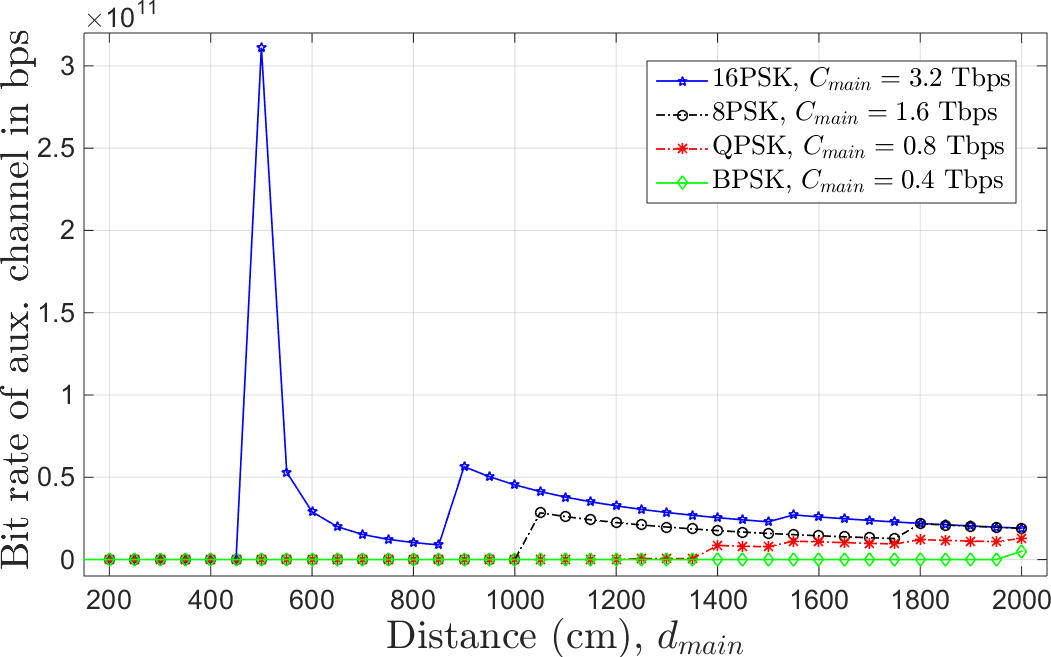}
\caption{Transmission rate of auxiliary channel $C_{aux}$ vs. transmission distance of main channel C, $d_{main}$, when $d_{aux}=500$ cm.}
\label{transratechannelC}
\vspace{-0.6cm}
\end{figure}

Fig. \ref{transratechannelC} presents the transmission rate of auxiliary channel, when the main THz channel is configured as a channel C. In the case of BPSK modulation, the auxiliary channel is only required, i.e., $C_{aux}=5.082 \cdot 10^9$ bps, when the transmission distance $d_{main}$ is larger than $2000cm$. Using QPSK, the transmission rate of auxiliary channel has a slight decrease with maximum at $1.290 \cdot 10^{10}$ bps and minimum at $0.054 \cdot 10^{10}$ bps, when the transmission distance $d_{main}$ is between $1250$ cm and $2000$ cm, respectively. When 8PSK is applied and the transmission distance $d_{main}$ is larger than or equal to $1050cm$, the transmission rate of auxiliary channel $C_{aux}$ gradually decreases from $2.839 \cdot 10^{10}$ bps to $1.275 \cdot 10^{10}$ bps, slightly increases up to $2.179 \cdot 10^{10}$ bps, if $d_{max}=1800$ cm, and then keeps decreasing up to $1.892 \cdot 10^{10}$ bps. For 16PSK, the auxiliary channel is necessary when the transmission distance $d_{main}$ is equal to $500cm$, whereby $d_{main}=d_{aux}$ and the transmission rate $C_{aux}$ achieves a peak of $3.111 \cdot 10^{11}$ bps. Depending on transmission distance of the main channel, the transmission rate of auxiliary channel vary in interval [$0.089 \cdot 10^{11},0.565 \cdot 10^{11}$] bps.

Considering Fig. \ref{transratechannelB} and Fig. \ref{transratechannelC}, the transmission rate of auxiliary channel increases with an increasing modulation level of the main THz channel. Moreover, the results confirm our statement in Eq. \eqref{dlimit} that the  lowest transmission rate of auxiliary channel can be reached when $d_{aux}<<d_{main}$, whereby the data from the main THz channel and the auxiliary channel will arrive at THz receiver at the same time reducing the buffering overhead.

\begin{figure}[t]
\centering
\includegraphics[width=1\columnwidth]{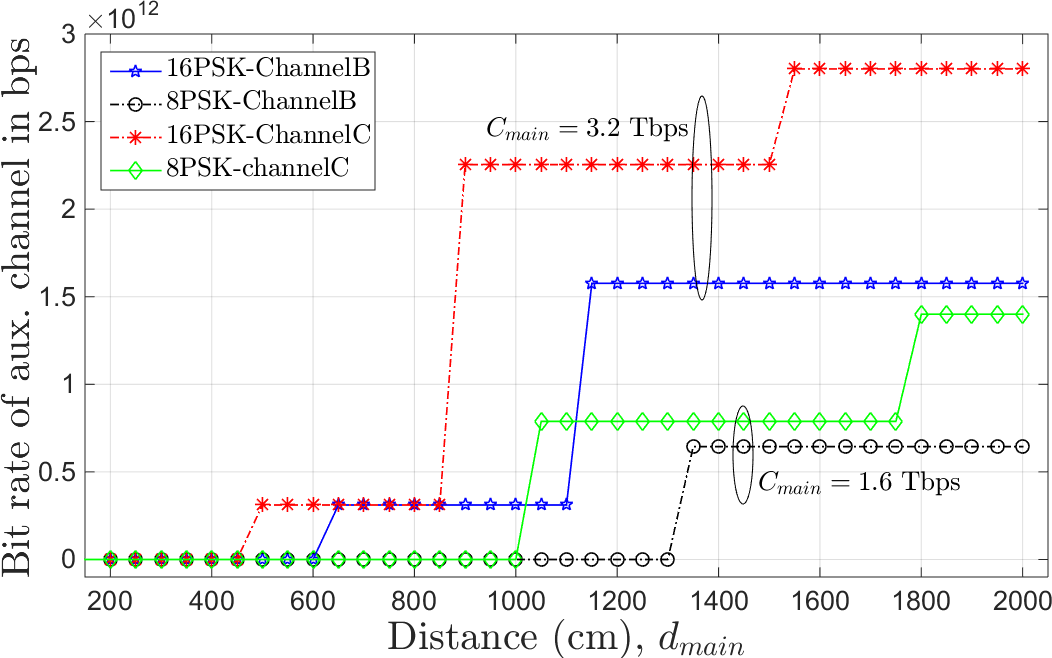}
\caption{Transmission rate of auxiliary channel $C_{aux}$ vs. transmission distance $d_{main}$ of main channels B and C, when $d_{aux}=d_{main}$.}
\label{transratechannelBC}
\vspace{-0.6cm}
\end{figure}

Next, we investigate the transmission rate of auxiliary channel $C_{aux}$, when the main and the auxiliary channels have the same transmission distance, i.e., $d_{main}=d_{aux}$. We compare two implementations of THz transmission system, where the main THz channel is either configured as channel B or as channel C. For this set of results, we consider only two high modulation formats, i.e., 8PSK and 16PSK resulting in the high transmission rates of the main THz channel defined as $1.6$ Tbps and $3.2$ Tbps, respectively.
Fig. \ref{transratechannelBC} shows a comparison of the transmission rates of auxiliary channels $C_{aux}$ as a function of the main channel configuration and transmission distance. When $d_{main}=d_{aux}$, $C_{aux}$ only depends on the amount of coding redundancy as can be derived from Eq. \eqref{rateaux}. Since the BER increases with increasing transmission distance $d_{main}$ and modulation level, the number of additional sRLNC redundancy for the erasure error correction increases requiring a high transmission rate of auxiliary channel $C_{aux}$. On the other hand, if the transmission distance $d_{main}$ slightly increases, the amount of required sRLNC redundancy and, thus, $C_{aux}$ keep unchanged. For example, using 16PSK with main channel C, the required transmission rate $C_{aux}=0.311 \cdot 10^{12}$ bps, when transmission distance vary from $d_{main}=500$ cm to $d_{main}=850$ cm. The reason is that the coding redundancy of $10$ symbols keeps constant for this distance interval requiring the constant transmission rate of auxiliary channel. In addition, $C_{aux}$ increases with an increasing modulation level. For the transmission distance $d_{main}=1500cm$ of the main channel B, we need an auxiliary channel with $C_{aux}=6.443 \cdot 10^{11}$ bps for 8PSK, while, for 16PSK, there is a need for the higher transmission rate up to $C_{aux}=1.577 \cdot 10^{12}$ bps. Based on results, we can conclude that $C_{aux}$ in THz transmission system with a main channel configured as channel B is often lower than that which is required to support the main channel configured as channel C. That is because the data sent over channel C suffer from more corrupted bits and, thus, need more additional redundant sRLNC symbols for successful data recovery at receiver. For example, with the same modulation format 8PSK, main channel B needs an auxiliry channel with transmission rate $C_{aux}=6.443 \cdot 10^{11}$ bps, while main channel C needs a supporting channel with transmission rate up to $C_{aux}=1.400 \cdot 10^{12}$ bps, when $d_{main}=d_{aux}=2000cm$.

\section{Discussion}
To make our proposed THz transmission system properly work, the auxiliary channel needs to have a very low BER/high SNR, while its transmission rate needs to be adapted to the parameters of the main THz channel. Since the main THz channel can reach transmission rate of up to 3.2 Tbps with a 16PSK modulation, 400 Gbytes of data per second will arrive at the receiver and have to be stored and processed, which is a challenge. When the data processing at receiver can not be made at line speed, i.e., at the speeds of THz signal, the amount of storage required will be prohibitively large. To minimize the data from the main THz channel that need to be stored for the coding redundancy from the auxiliary path, i.e., for symbol error correction, the data from the main and auxiliary channels have to simultaneously arrive at the receiver, which requires the proper selection of transmission rate and transmission distance of the auxiliary channel. From our evaluation of auxiliary channel with transmission distance $5$ m, the required transmission rate $C_{aux}$ can be between 200 Mbps and $3.2$ Tbps, which strongly depends on modulation format and transmission distance of the main channel. The highest transmission rate of auxiliary channel is required for the highest modulation format (16PSK) and a short transmission distance of the main channel and can be significantly reduced by using of low level modulation on the main channel. Generally, the transmission distance of auxiliary channel should be shorter than the transmission distance of the main channel, i.e.., to be able to utilize the lower transmission rate as can be seen from Eq. \eqref{dlimit}. Otherwise, either the transmission rate of auxiliary channel needs to be very high to timely deliver redundancy or the buffer size needs be larger. That rises a question about the suitable technology to implement and configure an auxiliary channel. 

If we use THz channel as the auxiliary, the transmission rate of up to 3.2 Tbps can be achieved as per investigated use case, whereby we need to ensure that the BER remains rather small, which is a challenge. To understand the tradeoff better, we created Table \ref{AuxCh} that can provide recommendations on auxiliary channel configurations. Depending on the distance between the sender and receiver, the auxiliary channel needs to use specific modulation format to reach $BER\to 0$. For instance, the maximum distance of $600 cm$ and $450 cm$ are supported by the modulation format of BPSK for channel B and channel C, respectively, or the maximum distance of $100 cm$ and $50 cm$ are supported by the modulation format of 16PSK for channel B and channel C, respectively. Of course there are other wireless or wireline technologies that can implement an auxiliary channel. When auxiliary channel requires transmission rate from Mbps area, the Radio Frequency (RF) or Free Space Optics (FSO) channel can be utilized as well. It should be noted however that one needs to consider that FSO and RF suffer from atmospheric effects such as fog and rain, respectively.

\begin{table}[t!]
  \centering
  \caption{Transmission distance and Modulation Format to reach $BER\to 0$ on Channel B and Channel C \cite{7444891}.}
  \label{AuxCh}
\resizebox{\columnwidth}{!}{%
\begingroup
\setlength{\tabcolsep}{0.7pt} 
\renewcommand{\arraystretch}{1.2} 
\begin{tabular}{ |c||c|c| }
 \hline
\multicolumn{1}{|c||}{\multirow{2}{*}{Modulation Format}} & \multicolumn{2}{|c|}{Distance (cm)}                                \\\cline{2-3} 
\multicolumn{1}{|c||}{}                                   & \multicolumn{1}{|c|}{Channel B ($660-695$ GHz)} & \multicolumn{1}{|c|}{Channel C ($855-890$ GHz)} \\\cline{2-3} 
 \hline
 16PSK   & $\leqslant$ $100$   &$\leqslant$ $50$\\
 8PSK&   $\leqslant$ $200$  & $\leqslant$ $150$   \\
 QPSK &$\leqslant$ $400$ & $\leqslant$ $300$\\
 BPSK    &$\leqslant$ $600$ & $\leqslant$ $450$\\
 \hline
\end{tabular} 
\endgroup
}\vspace{-0.6cm}
\end{table}

\section{Conclusion}\label{conc}
In this paper, we proposed and analyzed a THz transmission system that can address the challenges of coding complexity and throughput. Our idea is to first complement a generic low complexity FEC approach by a low complexity sRLNC. Since the overall throughput of THz transmission is a function of the coding redundancy, we proposed to send the native data and coding redundancy in parallel over two parallel channels. The provided analysis allows evaluation of the amount of required sRLNC redundancy and a code rate for a more reliable transmission over THz. The results confirmed, that the main channel can use a high-level modulation format and send over larger distances, e.g., $1100$ cm with 16PSK, in order to achieve a high transmission rate, when the lower transmission rate (at least $10$ Gbps) auxiliary channel delivers coding redundancy for the erasure error correction at receiver.

\section*{Acknowledgment}
This work was partially supported by the DFG Project Nr. JU2757/12-1, "Meteracom: Metrology for parallel THz communication channels."

\bibliographystyle{IEEEtran}
\bibliography{nc-rest}

\end{document}